\documentclass[12pt]{iopart}
\begin{document}

\title[On the criticality of frustrated spin systems with noncollinear
order]{On the criticality of frustrated spin systems with
noncollinear order}

\author{Yurij Holovatch$^{1,2}$, Dmytro
Ivaneyko$^{3}$ and Bertrand Delamotte$^{4}$
 }

\address{$^{1}$
Institute for Condensed Matter Physics and Ivan Franko National
University of Lviv, UA--79011 Lviv, Ukraine}

\address{$^{2}$
Institute f\"ur Theoretische Physik, Johannes Kepler Universit\"at
Linz, A-4040 Linz, Austria}

\address{$^{3}$
Ivan Franko National University of Lviv, UA--79005 Lviv, Ukraine}

\address{$^{4}$ Laboratoire de
Physique Th\'eorique et Hautes Energies,
 Universit\'es Paris VI-Pierre et Marie Curie -- Paris VII-Denis Diderot,
 75252 Paris Cedex 05, France}

\begin{abstract}
We analyze the universal features of the critical behaviour of
frustrated spin systems with noncollinear order. By means of the
field theoretical renormalization group approach, we study the
$3d$ model of a frustrated magnet and obtain pseudo-$\varepsilon$
expansions for its universal order parameter marginal dimensions.
These dimensions govern accessibility of the renormalization group
transformation fixed points, and, hence, define the scenario of
the phase transition.
\end{abstract}


\eads{\mailto{hol@icmp.lviv.ua},
\mailto{ivaneiko@ktf.franko.lviv.ua},
\mailto{delamotte@lpthe.jussieu.fr}} \maketitle


\section{Introduction} \label{I}

Remarkable progress achieved in the description of phase
transitions and critical phenomena due to application of the
renormalization group \cite{RGbooks} (RG) ideas leads sometimes to
the conclusion that all principal work in the field has been done,
especially if universal features of criticality are addressed.
This is certainly not true if one goes beyond a description of the
para-to-ferromagnetic phase transition in a standard $N$-vector
model \cite{Holovatch2004}, which belongs to the universality
class of the $O(N)$-symmetric $\phi^4$ theory. To give an example,
a study of realistic systems often calls for account of a
single-ion anisotropy \cite{reviews}, structural disorder
\cite{reviews} or frustrations \cite{note1} which might
essentially change the critical behaviour. An account of this new
physics still remains a challenging problem.

The theoretical RG description of the above mentioned systems
requires field theories of complicated symmetry with several
couplings. In theory, the critical point corresponds to the fixed
point of the RG transformation. An accessibility of the fixed
point, along with the (non-universal) initial conditions of the RG
flow is governed by the so-called marginal field dimensions. These
are universal and together with the critical exponents and
amplitude ratios constitute intrinsic features of criticality. It
is well established by now that the universal features of a 2nd
order phase transition in the $3d$ $N$-vector model are not
sensitive to the single-ion cubic anisotropy if $N<N_c^{\rm cub}$
\cite{reviews}. They are neither changed upon a weak quenched
dilution by a non-magnetic component if only \cite{reviews}
$N>N_c^{\rm dil}$. Currently, there exists a good agreement
between the numerical values of the marginal dimensions $N_c^{\rm
cub}$, $N_c^{\rm dil}$ calculated in numerous RG and MC studies
\cite{reviews,Folk00,margdil}. However, this is not the case for
the frustrated systems, where even more marginal dimensions have
been found and their numerical values are still under discussion
\cite{note1,Parruccini03}.

The problem we want to raise in this report concerns the critical
behaviour of  the $3d$ frustrated spin systems with noncollinear
order. The most common physical realization of such systems are
stacked triangular antiferromagnets (the examples are given by
${\rm CsMnBr_3}$, ${\rm CsNiCl_3}$, ${\rm CsMnI_3}$, ${\rm
CsCuCl_3}$, ${\rm VCl_2}$, ${\rm VBr_2}$) and helimagnets (${\rm
Ho}$, ${\rm Dy}$, ${\rm Tb}$, $\beta-{\rm MnO_2}$). In the former
case, the noncollinear order is caused by the frustrations due to
the triangular geometry of the underlying lattice, whereas in the
latter one it is due to the competition of ferro- and
antiferromagnetic interactions. Currently, there exists a large
literature devoted to the subject, which results from more than
twenty years long studies \cite{note1}. However, neither
experimentally nor theoretically has an unanimous conclusion been
drawn so far about the nature of the phase transition into the
ordered state in these systems. Important physical quantities
which are under discussion are the marginal dimensions of the
models. In particular, when the model is generalized to describe
$N$-component vectors \cite{Kawamura88}  (physical systems
mentioned above correspond to $N=2,3$), one finds a marginal
dimension $N_{c_3}$  below which the phase transition is of
first-order, whereas for  $N>N_{c_3}$ it is of second-order.
Several variants of the perturbative RG
expansions \cite{Pelissetto01,Sokolov03,Antonenko94,
Antonenko95,Pelissetto01a,Gracey02} and various truncations for
the Wilson-like non perturbative RG (NPRG) equations
\cite{Tissier03,Zumbach93,Tissier00,Tissier01} give different
numerical estimates for $N_{c_3}$ (see Table \ref{tab1}). However,
they all agree, that such a dimension (along with two marginal
dimensions more, $N_{c_2}$ and $N_{c_1}$, see below) exists. In
our study, we aim at performing a thorough analysis of these
marginal dimensions by means of the pseudo-$\varepsilon$ expansion
\cite{pseps}: the technique which is known to provide the most
accurate results in the 3d perturbative RG approach.

The rest of the paper is organized as follows: in the next section
we formulate the model we are interested in and obtain the
expansions for its marginal dimensions, section \ref{III} is
devoted to the numerical estimates on their basis, section
\ref{IV} gives conclusions and outlook.

\begin{table}[htp]
\caption {\label{tab1} Marginal dimension $N_{c_3}$ obtained
within different RG methods. See the text for details of the
methods.}
\begin{indented} \item[]
\begin{tabular}{|ccc|ll|ll|llll|}
\hline
 \multicolumn{3}{|c|} {fixed $3d$, resummed} &
 \multicolumn{2}{c|}
{$\sim \varepsilon^2$} &
 \multicolumn{2}{c|}
{$\sim 1/N$} &
 \multicolumn{4}{c|}
   {NPRG}
      \\
3 loops & \multicolumn{2}{c|} {6 loops} & & & & & & & &
      \\
\cite{Antonenko94}&
\cite{Pelissetto01}&
\cite{Sokolov03}&
  \cite{Antonenko95}&
    \cite{Pelissetto01a} &
 \cite{Pelissetto01a} &
 \cite{Gracey02} &
 \cite{Zumbach93} &
 \cite{Tissier00} &
 \cite{Tissier01} &
 \cite{Tissier03}
\\
      \hline
  3.91(1) &
  6 &
 6.4(4) &
 3.39    &
 5.3(2) &
 5 &
 3.24 &
 4.8 &
 4 &
 5 &
 5.1 \\
 \hline
\end{tabular}
\end{indented}
\end{table}


\section{The model and the pseudo-$\varepsilon$ expansion} \label{II}

An effective Hamiltonian of the model of  frustrated magnets with
an $N$-component order parameter reads \cite{Kawamura88}:
\begin{eqnarray}\nonumber
{\mathcal H} &=& \int{\rm d^d} x \Big\{
\frac{1}{2}[(\nabla\phi_1)^2+ (\nabla\phi_2)^2 + m_0^2
(\phi_1^2+\phi_2^2)] + \frac{u_0}{4!}[\phi_1^2+\phi_2^2]^2 +
\\&& \label{1}
\frac{v_0}{4!}[(\phi_1 \cdot \phi_2)^2- \phi_1^2\phi_2^2 ] \Big
\}.
\end{eqnarray}
In (\ref{1}), $m_0,u_0,v_0$ are bare mass and couplings and
$\phi_i\equiv\phi_i(x)$ are $N$-component vector fields,
representing the cosine and sine modes associated with the spin
ordering. The noncollinear (chiral) ordering occurs for
 $u,v\geq 0$.

We sketch the fixed points (FPs) picture retrieved in the previous
RG studies \cite{Kawamura88}. For the high field dimensions
$N>N_{c_3}$ four FPs exist: the Gaussian, $u=v=0$, $G$, the
Heisenberg $u\neq 0,v=0$, $O(2N)$ symmetric, $H$ (both unstable
for the space dimension $d=3$), and two nontrivial FPs
$u\neq0,v\neq0$ (with $u,v>0$), chiral and antichiral $C_+$ and
$C_-$. The FP $C_+$ is stable and governs the chiral 2nd order
phase transition. With a decrease of $N$, the FPs $C_+$ and $C_-$
merge at $N=N_{c_3}$ and disappear: only unstable FPs $G$ and $H$
are present for $N$ just below $N_{c_3}$. Note that Pelissetto et
al. and Calabrese et al. \cite{Pelissetto01,Sokolov03} have
claimed that, once resummed, the  six loop $\beta$-functions
obtained directly in $d=3$ exhibit a new root --- a new fixed
point --- below an extra marginal dimension of the field estimated
at $N\sim 5.7$. According to these authors, this fixed point is
neither analytically related to the Gaussian fixed point in $d=4$
nor to the fixed point found at large $N$ in $d=3$. It should
therefore be non perturbative, although it is found within the
perturbative framework.  Thus, they claim that below this new
marginal dimension, and in particular for $N=2,3$, the transition
is again of second order. Note also that this scenario disagrees
with the results obtained within the NPRG approach for which the
transition is (very weakly) of first order for $N<N_{c_3}$. We
will return to this point at the end of our report. As $N$ is
further decreased, the nontrivial FPs $C_+$ and $C_-$, existing
above $N_{c_3}$ reappear  at $N\leq N_{c_2}$, but now in the
$u>0,v< 0$ quadrant and thus do not describe the chiral phase
(they both have complex coordinates for $N$ between $N_{c_3}$ and
$N_{c_2}$). Finally at $N=N_{c_1}$ one of the nontrivial FPs
merges with the FP $H$ and, with further decrease of $N$, passes
to the quadrant $u> 0,v> 0$, still remaining unstable. The above
picture is supported both by the perturbative RG approach
($\varepsilon$-expansion accompanied by subsequent resummation
\cite{Antonenko95} or by a conjecture about the series behaviour
\cite{Pelissetto01a}, resummed expansion in terms of renormalized
couplings in a $3d$ RG scheme
\cite{Pelissetto01,Sokolov03,Antonenko94}, $1/N$ expansion
\cite{Pelissetto01a,Gracey02}) and the non-perturbative RG
\cite{Tissier03,Zumbach93,Tissier00,Tissier01}.

Discrepancies between the values of the marginal dimensions
$N_{c_i}$ obtained so far within the perturbative RG (see Table
\ref{tab1}  for $N_{c_3}$) to a great extent are because the
series used for their analysis were rather short
\cite{Antonenko94,Antonenko95,Pelissetto01a,Gracey02} and in
general are known to be asymptotic at best \cite{RGbooks}. So it
is very desirable to perform an estimate of the marginal
dimensions on the basis of the expansions which, on the one hand
would be of the highest order and, on the other hand, would
possess better convergence properties. As we show below, these two
goals are reached by applying the pseudo-$\varepsilon$ expansion
to the six-loop $d=3$ RG $\beta$-functions obtained for the
effective Hamiltonian (\ref{1}) in the Ref. \cite{Pelissetto01}.

The method consists in introducing an auxiliary parameter ($\tau$)
into the $\beta$-functions which allows to separate contributions
to the FPs from the loop integrals of different order
\cite{pseps}. This is achieved by multiplying a zero-loop term by
$\tau$ and obtaining FP coordinates and, subsequently, all FP
quantities of the theory as series in $\tau$ with final
substitution $\tau=1$. Starting from the six-loop $d=3$ RG
$\beta$-functions of Ref. \cite{Pelissetto01} we get the
expansions for the fixed point coordinates and derive  for the
marginal dimensions:
\begin{eqnarray} \nonumber
N_{c_3} &=& 21.797959-15.620635\,\tau+ 0.262060\,{\tau}^{2}-
0.150930\,{\tau}^{3}-
\\ && \label{2}
 0.039165\,{\tau}^{4} - 0.029721\,{\tau}^{5},
\\ \nonumber
 N_{c_2} &=&  2.202041-
0.379365\,\tau+0.202166\,{\tau}^{2}- 0.084951\,{\tau}^{3}+
\\ && \label{3}
0.092744\,{\tau}^{4}-
 0.097961\,{\tau}^{5}, \\
\nonumber
 N_{c_1}&=& 2.0- 0.666667\,\tau+ 0.145212\,{\tau}^{2}-
0.094836\,{\tau}^{3}+
\\ && \label{4}
0.099757\,{\tau}^{4} -
 0.112325\,{\tau}^{5}.
\end{eqnarray}

Expansions (\ref{2})--(\ref{4}), derived directly at $d=3$, can be
compared with the corresponding $\varepsilon=4-d$-expansions
\cite{Antonenko95}:
\begin{eqnarray}\label{5}
N_{c_3}&=&21.80-23.43\varepsilon+7.088\varepsilon^2,\\
 \label{6}
N_{c_2}&=&2.202-0.569\varepsilon+0.989\varepsilon^2, \\
 \label{7}
N_{c_1}&=&2-\varepsilon+1.294\varepsilon^2.
\end{eqnarray}
Formulas (\ref{2})--(\ref{4}) take into account three orders of
perturbation theory more than the highest available
$\varepsilon=4-d$-expansions (\ref{5})--(\ref{7}). Moreover,
comparing (\ref{2})--(\ref{4}) and (\ref{5})--(\ref{7}) one sees
that the expansion coefficients in the pseudo-$\varepsilon$ series
decay much faster and one may expect to get more convergent
results on their basis. And indeed this is the case as we will see
in the next section.


\section{Numerical estimates of the marginal dimensions} \label{III}

The field theoretic RG expansions are known to have zero radius of
convergence and different resummation techniques are used to make
numerical estimates on their basis \cite{RGbooks}. Here, we make
use of the Pad\'e-analysis \cite{Baker81} to make an analytic
continuation of the expansions for $\tau=1$. On the one hand
already this simple technique allows us to show essential features
of the pseudo-$\varepsilon$ expansion behaviour, on the other hand
it allows to determine numerical values of the marginal dimensions
with a sufficient accuracy. The results for the
pseudo-$\varepsilon$ expansion series (\ref{2})--(\ref{4}) are
given below in the form of Pad\'e-tables. There, a result of an
$[M/N]$ Pad\'e approximant is represented as an element of a
matrix with usual notation, e.g. the first row gives results of
the mere summation of the series:

\begin{eqnarray}
 \label{8}
N_{c_3}&=&\left[ \begin {array}{cccccc}
                     21.798& 6.177&6.439&6.288&6.249&6.220
\\\noalign{\medskip} 12.698&6.435& 6.344&6.236&\frac{6.126}{1.318}&
\\\noalign{\medskip} 9.827&6.290&6.230&\frac{6.182}{1.751}&&
\\\noalign{\medskip} 8.463&6.247&\frac{6.155}{1.453}&&&
\\\noalign{\medskip} 7.695&6.217&&&&
\\\noalign{\medskip} 7.220&&&&&
\end {array}\right]
 \\ \label{9}
 N_{c_2}&=&\left[ \begin {array}{cccccc}
2.202&1.823&2.025& 1.940& 2.033&1.935\\\noalign{\medskip}
1.878&1.955&1.965&1.984&1.985&\\\noalign{\medskip}
1.984&1.966&\frac{1.948}{0.385}&1.985&&\\\noalign{\medskip}
1.962&1.977&1.988&&&\\\noalign{\medskip}
\frac{2.012}{2.586}&1.986&&&&\\\noalign{\medskip} 1.960&&&&&
\end {array}\right]
 \\ \label{10}
 N_{c_1}&=& \left[
\begin {array}{cccccc} 2.0& 1.333&
1.479&1.384&1.483&1.371\\\noalign{\medskip} 1.500& 1.453&
1.421&1.432&1.431&\\\noalign{\medskip}
1.458&\frac{1.032}{1.105}&1.431&1.431&&\\\noalign{\medskip}
1.421&1.436&1.431&&&\\\noalign{\medskip} 1.446&1.432&&&&
\\\noalign{\medskip} 1.415&&&&&
\end{array}
\right]
\end{eqnarray}
Results shown by fractions indicate, that the Pad\'e approximant
contained a pole for $\tau$ close to 1 (the denominator shows the
value of $\tau$ for which the pole is obtained). These tables can
be compared with the analogous tables for the
$\varepsilon$-expansions (\ref{5})--(\ref{7}):
\begin{eqnarray}\label{11}
N_{c_3}&=&  \left[
 \begin{array}{ccc}

 21.80        & -1.630        & 5.458  \\
 10.507       & 3.812         &        \\
 7.505        &               &
\end{array} \right],
 \\ \label{12}
N_{c_2}&=&  \left[
 \begin{array}{cccc}

 2.202      &1.633          & 2.622  \\
1.750       & 1.994         &        \\ 2.514       & &

\end{array} \right],
 \\ \label{13}
  N_{c_1}&=&   \left[
 \begin{array}{cccc}

 2        &1        & 2.294  \\
 1.333    & 1.564   &        \\
 1.813   &         &

\end{array} \right].
\end{eqnarray}
One certainly sees that the convergence properties of the
pseudo-$\varepsilon$ expansion are better in comparison with the
$\varepsilon$-expansion (cf. the convergence of the results along
the main diagonal and those parallel to it: there the Pad\'e
analysis is known to provide the most reliable data
\cite{Baker81}). One more feature of the expansions for $N_{c_i}$
is evident when one compares tables (\ref{8})--(\ref{10}): whereas
the central elements of the table (\ref{10}) give a firm estimate
for $N_{c_1}$: $[2/2]=[3/2]=[2/3]=1.431$, such a stable behaviour
is not found in the corresponding Pad\'e tables (\ref{8}),
(\ref{9}) for $N_{c_3}$, $N_{c_2}$. Obviously, this different
behaviour is connected with the different origin of the marginal
dimensions $N_{c_1}$ from the one side, and $N_{c_3}$, $N_{c_2}$
from the other side. Indeed, the dimension $N_{c_1}$ corresponds
to merging of the non-trivial and Heisenberg FPs after which the
non-trivial FP continuously passes to the other quadrant of the
$u-v$ plane whereas dimensions $N_{c_3}$, $N_{c_2}$ correspond to
the coalesce and {\em disappearance} of two non-trivial FPs (see
discussion at the beginning of section \ref{II}).

To make the numerical estimates on the base of Pad\'e-tables
(\ref{8})--(\ref{10}) we proceed as follows. For $N_{c_3}$ we take
on the main diagonal the highest Pad\'e approximant with an
estimate $[2/2]=6.23$ and suppose that the deviations from an
account of higher-order terms will not exceed the difference
$[2/2]-[1/1]=0.21$. For $N_{c_2}$ we take the highest obtained
estimates  $[3/2]=[2/3]=1.99$ considering a confidence interval as
$[3/2]-[2/2]=0.04$. Subsequently, for $N_{c_1}$ the central value
is given by $[3/2]=[2/3]=[2/2]=1.43$ with a confidence interval
$[2/2]-[1/1]=0.02$. Finally, we get for the marginal dimensions:
 \begin{equation} \label{14}
 N_{c_3}=6.23(21), \hspace{1em}
 N_{c_2}=1.99(4), \hspace{1em}
 N_{c_1}=1.43(2).
 \end{equation}
The above estimates include within the error bars {\em all}
elements of corresponding Pad\'e-tables except of the inverse
approximants $[0/N]$ (and the approximant $[5/0]$ of (\ref{9}))
and therefore the confidence intervals in (\ref{14}) are rather
overestimated.

Comparison of our estimate for $N_{c_3}$ with the perturbative RG
data of Table \ref{tab1} supports recent estimates
\cite{Pelissetto01,Sokolov03} $N_{c_3}\sim 6$. We also suggest
that essential difference between this estimate and the numbers
obtained within $\varepsilon$- and $1/N$-expansions
\cite{Antonenko95,Pelissetto01a,Gracey02} is because the last have
not been estimated with comparative accuracy which was caused in
particular by shortness of corresponding series \cite{note2}.
Available so far estimates of $N_{c_2}$ are due to the resummation
of three-loop massive RG expansions \cite{Antonenko94} and of the
$\varepsilon^2$ expansion \cite{Antonenko95} (\ref{6}):
$N_{c_2}=1.96$ and $N_{c_2}=2.03(1)$, correspondingly
\cite{note2}. Together with the symmetry arguments
\cite{Antonenko94} providing $N_{c_2}>2$ our estimate suggests
that the value of $N_{c_2}$ should be located very close to 2. In
particular this means that corresponding scenario of appearance of
the pair of non-trivial FPs which is governed by this marginal
dimensions might not be found in numerical calculations for $N=2$.
Dimension $N_{c_1}$ has its counterpart \cite{Antonenko94} as the
marginal dimension of the $N$-vector model with a single-ion cubic
anisotropy: $N_{c_1}=N_c^{\rm cub}/2$, see section \ref{I}. The
last has been estimated by different methods. In particular the
Pad\'e-Borel resummation of the series (\ref{4}) gives
\cite{Folk00} the number coherent with the other data
\cite{reviews} $N_{c_1}=N_c^{\rm cub}/2=1.431(3)$. Comparison of
this number with our estimate (\ref{14}) based on much less
elaborated technique supports the reliability of chosen here
scheme.


\section{Conclusions} \label{IV}

The numerical values of the  marginal dimensions that we obtain
represent clearly an improvement of the preceding determinations
performed both by the perturbative methods and the NPRG one. Once
again the pseudo-$\varepsilon$ expansion turns out to be very
accurate and constitutes probably a new way to analyze the
critical behaviour of $3d$ frustrated magnets. However, the
principal question about the order of the phase transition in
these systems for $N=2,3$ still remains open. Obviously, our
studies are in coherence with the FP picture of the NPRG and
perturbative RG approaches, where no stable FP are found in the
region $N_{c_3}>N>N_{c_2}$ (however the difference between
numerical value of $N_{c_3}$ and typical numerical values found in
the NPRG studies \cite{Tissier03,Zumbach93,Tissier00,Tissier01}
calls for a more detailed analysis). Nevertheless, the existence
of the FP found recently in Refs. \cite{Pelissetto01,Sokolov03}
and claimed to describe criticality of these systems can neither
be supported nor rejected by our perturbative approach. We think
that one of possible ways to shed light on this problem is to try
to follow an evolution of this FP with change of $d$ in order to
understand its origin at the upper critical dimension $d=4$.

\ack

This work was supported in part by the Austrian Fonds zur
F\"orderung der wissenschaftlichen Forschung, project
No.16574-PHY.

\vspace{0.9cm}

{\small {\it Note added.} After the completion of this work
\cite{Delamotte03}, we learned about $\varepsilon^4$-results for
$N_{c_3},N_{c_2}$ and a six-loop pseudo-$\varepsilon$ result for
$N_{c_3}$ \cite{Calabrese03}. The five-loop
$\varepsilon$-expansion improves the $\varepsilon^2$ data for
$N_{c_3}=6.1(6)$ but leads to the unphysical conclusion
$N_{c_2}=1.968(1)<2$. The pseudo-$\varepsilon$ expansion of Ref.
\cite{Calabrese03} for $N_{c_3}$ coincide with our formula
(\ref{2}), but we report higher numerical accuracy.}

\end{document}